\documentclass{jpp}
\usepackage{graphicx}
\usepackage{epstopdf, epsfig}

 \newcommand{\bfE}{\mathbf{E}}

\newcommand{\bfB}{\mathbf{B}}

\newcommand{\bfJ}{\mathbf{J}}

\newcommand{\bfv}{\mathbf{v}}

\newcommand{\alf}{{Alfv\'en }}
\newcommand{\bfbnorm}{\mathbf{b}}

\shorttitle{Non-linear Waves and Instabilities  in  Reconnection Outflows}

\title{Non-linear Waves and Instabilities Leading to Secondary Reconnection in  Reconnection Outflows}

\author{Giovanni Lapenta\aff{1,2} 
  \corresp{\email{giovanni.lapenta@kuleuven.be}}
\shortauthor{G. Lapenta, F. Pucci, V. Olshevsky et al.}, Francesco Pucci\aff{1}, Vyacheslav Olshevsky\aff{1},  Sergio Servidio\aff{3}, Luca Sorriso-Valvo\aff{4},      David L. Newman\aff{5} \and Martin Goldman\aff{5}}
\affiliation{\aff{1}Depatment of Mathematics, KULeuven University, Leuven, Belgium
\aff{2} Space Science Institute, Boulder, USA
\aff{3}Dipartimento di Fisica, Universit\`a della Calabria, Cosenza, Italy
\aff{4}Nanotec-CNR, U.O.S. Cosenza, Arcavacata di Rende, Italy
\aff{5}University of Colorado, Boulder, USA

}
\begin{document}

\maketitle

\begin{abstract}
Reconnection outflows are  regions of intense recent scrutiny, from in situ observations and from simulations. These regions are host to a variety of instabilities and intense energy exchanges, often even superior to the main reconnection site. We report here a number of results drawn from investigation of simulations. First, the outflows are observed to become unstable to drift instabilities. Second, these instabilities lead to the formation of secondary reconnection sites. Third, the secondary processes are responsible for large energy exchanges and particle energization. Finally, the particle distribution function are  modified to become non-Maxwellian and include multiple interpenetrating populations.
\end{abstract}

\section{Introduction}

The research of the last two decades has shown that kinetic reconnection is a fast process that develops on \alf time-scales \citep{biskamp}. This result is a spectacular success for kinetic modelling \citep{birnGEM}, now confirmed in situ by the Magnetospheric Multiscale Mission \citep{burch2016electron}. However, fast kinetic reconnection is not the solution to all problems in reconnection: fast kinetic reconnection has thus far been observed and modelled only in localised regions. Instead, in many astrophysical and laboratory systems, large amounts of energy are converted over large domains.  How can we bring fast kinetic reconnection to large scales?

A possible scenario to reach large energy conversion rates on system scales is to imagine a situation where the initiation of reconnection is followed by a chain reaction of more and more secondary reconnection sites \citep{bulanov1979tearing,loureiro2007instability,lapenta08,tenerani2016ideally}. Under these conditions, reconnection tends to become chaotic with many reconnection sites being spawned by instability and reabsorbed by island coalescence, leading to fast reconnection \citep{bhattacharjee09,skender,pucci2013reconnection,huang2017plasmoid}.

Three dimensional reconnection is accompanied by many more instabilities than just the formation of secondary islands in the primary reconnection site seen in two dimensional reconnection: the reconnection inflow\citep{daughton2011role} and the reconnection outflow~\citep{lapenta2015secondary} host instabilities that lead to secondary reconnection. The first mechanism is primarily present in reconnection separatrices in the case of strong guide fields \citep{lapenta2016reconnection}, while the latter is present at all guide fields \citep{lapenta2014separatrices}. 

Outflows from reconnection are rich in free energy that can drive instabilities. Among the possibilities we consider here:
\begin{itemize}
\item Velocity shears around the outflow jet that can drive Kelvin-Helmholtz instability \citep{lottermoser1998ion}.
\item Density and temperature gradients at the front formed by the outflowing jet interacting with the ambient plasma leads to drift-type instabilities \citep{divin2015evolution}.
\item Unfavourable curvature of field lines between the separatrices in the outflow region can lead to interchange (Rayleigh-Taylor-type) instabilities \citep{nakamura2002interchange,guzdar2010simple,lapenta2011self}.
\item Flux ropes in the outflows may be kink unstable \citep{kruskal1958instability,shafranov1957equilibrium}.
\item Additional instabilities are caused by phase-space features such as anisotropies leading to whistler waves and beams leading to streaming instabilities \citep{goldman2016can}.
\end{itemize}
All these instabilities can cause strong deformation of the flow, leading   possibly to turbulence \citep{pucci2017properties}, energy exchange \citep{lapenta2016energy} and secondary reconnection~\citep{lapenta2015secondary}.

The 3D scenario for large scale turbulence is than one where reconnection might lead to a chain-reaction type of sequence of events. Reconnection is initiated at one location but the instabilities associated with the flows and the other sources of free energy induced by reconnection lead to the formation of secondary reconnection sites. While not yet observed in simulation, this scenario on large scales (not yet accessible to simulation) can then progress in successive generation of tertiary and further reconnection sites, filling macroscopic domains.

Below, we organize our material as follows. Section 2 reports the type of simulations we use to analyze the reconnection outflows and  the instabilities developing there. Section 3 investigate the fluctuation spectrum produced in the outflow. Section 4 discusses how the fluctuations interact with the particles energizing them. Conclusions and future directions are outlined in Sect. 5.

\section{Development of outflow instabilities and secondary reconnection}

In order to study the properties of  outflow instabilities and secondary reconnection, we use particle-in-cell numerical simulations.

We consider here the same run previously considered in \citet{lapenta2015secondary}. The system is initialized with a Harris equilibrium \citep{Harris1962} 
\begin{equation}
{\bf B} = B_{0x}\tanh{(y/\delta)}{\bf e_x} + B_{0z}{\bf e_z},\;\;\; n=n_{0b}+\frac{n_0}{\cosh^2(y/\delta)}. 
\end{equation}
uniquely specified by the mass ratio $m_i/m_e= 256$,  the temperature ratio $T_i/T_e= 5$ and $v_{th,e}/c=0.045$. We set the density of the uniform plasma background to $n_{0b} = n_0/10$ and the value of the guide field to $B_{0z}/B_{0x}=1/10$. The evolution is followed using the fully electromagnetic and fully kinetic iPic3D code \citep{iPIC3D} that treats both electrons and ions as particles. Details are provided in \citet{lapenta2015secondary}. We use coordinates where $x$ is along the initial magnetic field, $y$ is along the initial gradients, and $z$ is along the initial current. Open boundary conditions are imposed in the $x$ and $y$ direction and periodicity is imposed along $z$. 
We consider a 3D box of shape $[40.0,15.0,10.0]d_i$, where $d_i$ is the ion inertial length, which is resolved by a
cartesian grid of $[512,192,128]$ cells, each one populated with 125 particles. The spatial resolution is $\Delta x = 1.25d_e$, where $d_e$ is the electron inertial length, and the time step is $\Delta t=\pi/10 \omega_{ce}^{-1}$, where $\omega_{ce}$ is the electron gyro-frequency.  

\begin{figure}
  \centerline{\includegraphics[width=\columnwidth]{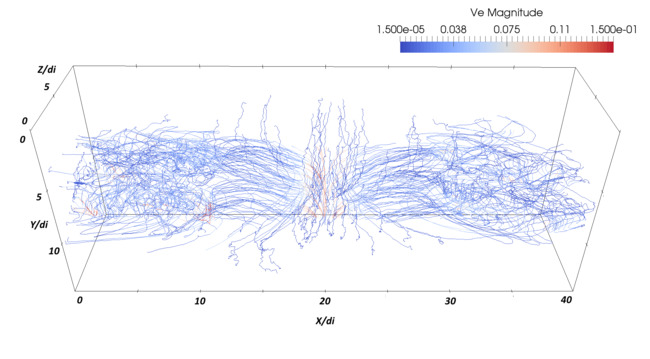}}
  \caption{Visualization of the electron flow around a reconnection site. We report streamlines of the first order moment of the electron distribution (the electron flow velocity) coloured by the intensity of the local electron speed (normalized to the speed of light). }
\label{fig:front1}
\end{figure}

Reconnection is initialised in the centre with an initial x-shaped perturbation that leads to the formation of a central x-line. A reconnection site develops with plasma accelerated towards the reconnection region and expelled out of it. The electron flow pattern in the fully developed non-linear stage is shown in Figure~\ref{fig:front1}. The electrons are first attracted toward the central x-line where the z-directed reconnection electric field accelerates them to high speed. The Lorentz force then deflects the particles towards the outflow. In this region, the system presents a remarkable invariance along $z$, resembling the same physics of two dimensional fast kinetic reconnection.

In the outflow, however, the electron flow pattern becomes distorted and meanders about, eventually passing downstream away of the reconnection region. In this region the electron flow becomes more turbulent.  

The region of electron meandering corresponds to the front formed by the interaction of the outflowing plasma with the surrounding media. At the front, an effect similar to that of a snowplow pushes the plasma outward. A form forms where at least three of the mechanisms mentioned above are present: the field lines wrap around the front gaining unfavourbale curvature that can lead to interchange-type instabilities, the density gradient is unstable to drift modes and the distribution function becomes severely non-maxwellian leading to microinstabilities.  
%
%

\begin{figure}
  \centerline{\includegraphics[width=.5\columnwidth]{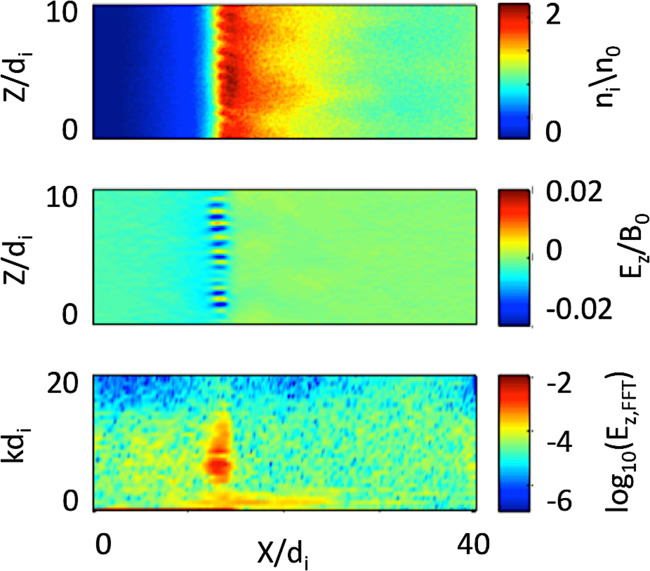}}
  \caption{Early stages of the instability at the front, at time $\omega_{ci}^{-1}=15$. The panels show from top to bottom:  ion density (a),  $z$-component of the electric field (b) and the Fourier spectrum in $k_z$ of the perturbation of the electric field $E_z$.}
\label{fig:LH2}
\end{figure}

Figure~\ref{fig:LH2} shows the state of the front after the instability starts to develop. The density  (panel a) becomes rippled by a mode that presents a strong perturbation of the $E_z$ (panel b). When the mode structure of these fluctuations is Fourier analysed, the resulting spectrum in $k_z$ is reported in panel c. The observed features are characteristic of a drift mode in the lower-hybrid range. 

\begin{figure}
  \centerline{\includegraphics[width=.7\columnwidth]{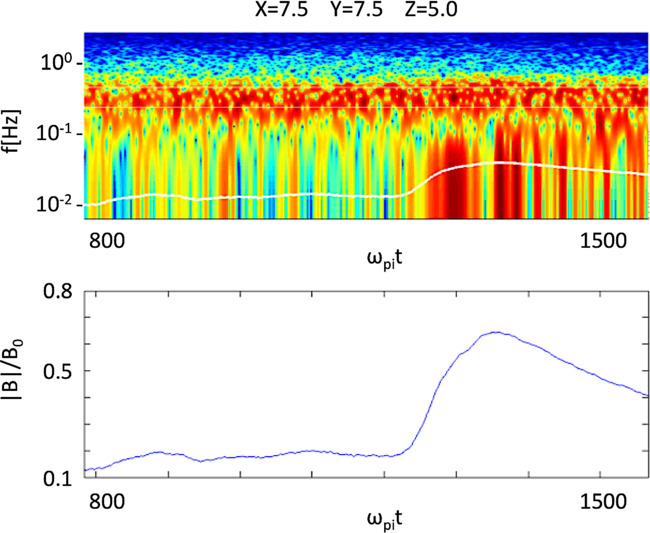}}
  \caption{Signal from a virtual probe embedded in the simulation at $x/d_i=7.54$, $y/d_i=7.54$, $z/d_i=5.04$. The top panel shows the spectrogram of the $E_z$ signal measured. To guide the eye the local lower hybrid frequency is indicated by a white line. The bottom panel shows the  magnetic field intensity measured by the virtual probes at the different times.}
\label{fig:LH1}
\end{figure}

The identification of the instability as having primarily the nature of a lower-hybrid drift instability (LHDI) is confirmed by the temporal spectrum measured at a fixed point reached by the front \citep{divin2015evolution,divin2015lower}. A spectrogram, obtained with standard windowing methods similar to those used in on-board real space probes, is reported:  the observed frequency spectrum is reported at different times. The lower panel reports the corresponding observed local magnetic field intensity.  When the front arrives, an intense signal in the lower hybrid range is measured. 

\begin{figure}
  \centerline{\includegraphics[width=\columnwidth]{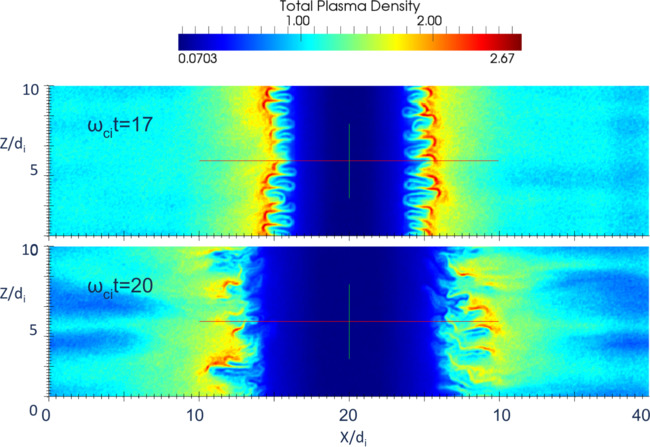}}
  \caption{Density at the front for two different times.}
\label{fig:front2}
\end{figure}

As the evolution is continued, the ripples in the front become more intense and start to interact leading to conditions where magnetic field of opposite polarity is brought in contact promoting secondary reconnection. 
Figure~\ref{fig:front2} shows the front at two consecutive times: at later times, the "fingers" formed in the front tend to interact and coalesce \citep{vapirev2013formation}.

\citet{lapenta2015secondary} analysed several indicators to detect positively secondary reconnection sites: direct analysis of field line connectivity, energy conversion in the electron frame ($\bfJ \cdot (\bfE +\bfv_{e} \times \bfB)$), electron agyrotropy, slippage ($\bfv_{e\perp} -\bfE \times \bfB/B^2$), topological  measure of field line breakage~\citep{hesse1988theoretical,biskamp} ($\bfbnorm \times \nabla \times (E_{||}\bfbnorm)$), where $\bfbnorm$ is the unit vector along $\bfB$ and $E_{||}$ is the parallel component of the non ideal part of  Ohm's law \citep{biskamp}, normalized as $eE_{||}/m_i c\omega_{pi}$. 

\begin{figure}
  \centerline{\includegraphics[width=\columnwidth]{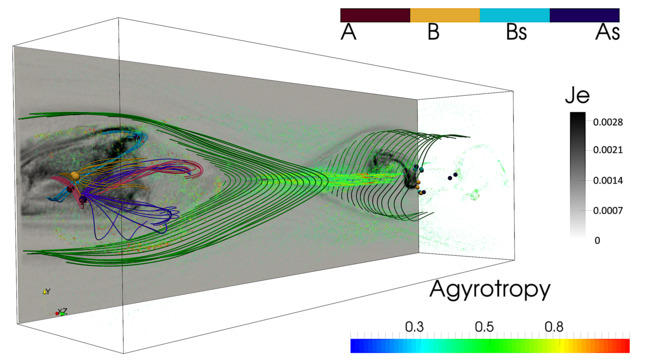}}
  \caption{Combination of different measures at the same time: a vertical cut of the electron current intensity (grayscale); false-colour volume rendering of agyrotropy; magnetic null points (colour spheres) coloured according to their topological type. Selected field lines reconnected once at the primary site are shown in green, while secondary reconnected lines near the nulls are shown in purple, pink, orange and light blue.}
\label{fig:nulls_slavik}
\end{figure}

A specific orientation of the magnetic field which allows for the field annihilation and energy release, is an important indicator of magnetic reconnection. In the classical two dimensional picture magnetic field lines of opposite direction approach each other and form an X-point, which, extended to 3D, becomes an X-line denoted by the strong Z-aligned current in our simulation. This sort of magnetic reconnection, however, does not require the field to become exactly zero (hence, no magnetic nulls are formed) on the reconnection site. We use the technique based on the topological degree method \citep{Greene:1992} as described in \citet{Olshevsky:2016ApJ} to locate and classify magnetic nulls. Indeed, in the simulation reported here no magnetic nulls are present in the central current sheet as summarized in Figure~\ref{fig:nulls_slavik}. However, the diffusion region around the X line is characterized by strong electron agyrotropy $A=\left(P_{e,\perp 1}-P_{e,\perp 2}\right)/P_{e,\perp 1}-\left(P_{e,\perp 2}\right)$ that is shown with volume rendering. No strong energy conversion is associated with the `main' reconnection X-line.

In contrast, in the reconnection outflow a number of magnetic nulls form which are depicted by colour spheres in Figure~\ref{fig:nulls_slavik}. The colour denotes magnetic null's topological type: A and B (red and orange) are the three-dimensional extensions of the X-points called radial nulls; while As and Bs (light blue and blue) represent magnetic islands or magnetic flux ropes. Both radial and spiral nulls are present in the outflows, however the number of spiral ones is larger. Magnetic field lines in the vicinity of the null points in the left outflow are shown with the corresponding colours. A pair of spiral nulls is formed in a swirl of the light blue magnetic field lines, probably, driven by a shear instability. This null pair is embedded in the region of strong energy conversion (see Figure~\ref{fig:nulls_slavik}). Other nulls in this outflow are on the interfaces of magnetic fields of different polarities characterized by complex field patterns resembling an X pattern (orange) and merging into flux ropes (purple and pink). Recent observations \citep{Fu:etal:2017} provide a strong evidence that intermittent energy conversion in the reconnection outflows is associated with the spiral magnetic nulls and twisted magnetic fields.

The picture provides an indication of the scenario described in the introduction: the initial reconnection site located at the centre of the box and forming a X-line produces two outflows that become unstable and produce in turn secondary reconnection sites. In the process the plasma becomes effectively turbulent and a large fraction of the energy is converted to particle heat at these unstable fronts, rather than at the central X-line.

\section{Development of intermittent turbulence in reconnection outflows}

Above we have observed how reconnection tends to become visually turbulent. But is turbulence real? 
In a recent paper \citep{pucci2017properties}, the properties of electric and magnetic fluctuations that are produced by magnetic reconnection have been analysed. 
Because of the inhomogeneous background it is important to first establish the anisotropy level and in general the 3D properties of turbulence. Analysis of the autocorrelation function of the magnetic field fluctuations have shown that the turbulence that develops in the reconnection jets is anisotropic. 
In particular, magnetic vortexes are elongated in the direction of the background magnetic field, namely $x$, with a second smaller anisotropy in the $yz$ plane. 
The second anisotropy becomes negligible for smaller scales and isotropy is recovered in the $(k_y,k_z)$ plane for $k_{yz} > 1.5$, with $k_{yz}=\sqrt{k_y^2+k_z^2}$. 
This allows to reduce the 3D spatial spectra to 1D isotropic spectra computed in $(k_y,k_z)$ plane and integrated in $k_x$. 
The results of this computation shows magnetic and electric spectra with a clear power law in the sub-ion range $1.5 < k_{yz}d_i < 15$. As observed in space plasmas \citep{eastwood2009observations} the magnetic and electric spectra departs from each other at around $kd_i \sim 1$, the electric one proceeding with a spectral slope of $\sim 1$ and the magnetic one with a slope of $-8/3$.   
Recently \citet{matteini2016electric}, following simple dimensional arguments, have interpreted this phenomenon as due to the dominance of the Hall-effect at small scales. 
It is worth remarking how this interpretation still holds in such an anisotropic and inhomogeneous system, where spectra need to be carefully extracted removing large-scale background profiles and border effects.

\begin{figure}
\includegraphics[width=\textwidth]{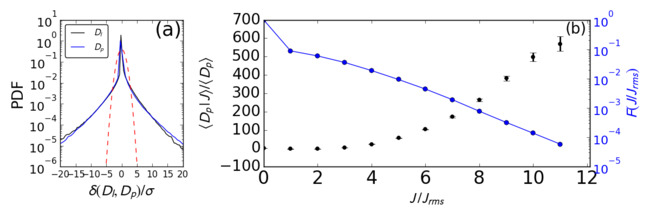}
\caption{PDFs of $D_l$ and $D_i$ (a). Red dashed lines 
represent the normalized Gaussian curve. Mean $D_i$ conditioned on local current density 
thresholds and (right axis) fraction $F$ of the full box data used to compute the averages (b).}
\label{fig:Int_fb}
\end{figure}

Turbulence is responsible for the transfer of energy from fields to particles. In this work we show that this energy exchange do not take place homogeneously in the reconnection events but is located in small regions in the reconnection outflows where the energy transfer is very intense. In order to quantify the energy exchange we introduce the two dissipation proxies $D_l= {\bf J} \cdot {\bf E}$ and $D_i = {\bf J} \cdot ({\bf E} + {\bf v}_i \times {\bf B})$ \citep{zenitani2011new}, where ${\bf J}$ is the total current, ${\bf E}$ is the electric field, ${\bf v}_i$ is the ion fluid velocity, and ${\bf B}$ is the magnetic field. 
In panel (a) of Figure~\ref{fig:Int_fb} the Probability Density Functions (PDFs) of $\delta D_l = D_l - \langle D_l \rangle_{x,y,z} $ 
and  $\delta D_i = D_i - \langle D_i \rangle_{x,y,z} $ are plotted, where $\langle \rangle_{x,y,z}$ means average along the three axes.
The two PDFs are compared with the normalized Gaussian distribution (plotted in dashed-red line). They 
strongly depart from Gaussian distributions, presenting instead high tails up to several standard 
deviations $\sigma$. 
In panel (b), the average $D_i$ conditioned to a threshold current density is shown. The plot is constructed as follows: a threshold in the current density magnitude is considered and the average of $D_i$ is computed using all those points in the domain where the value of the current is bigger than the fixed threshold. This average is then normalized to the average of $D_i$ on all points, which gives by definition $\langle D_i | J=0 \rangle/\langle D_i \rangle = 1$.
The black points in the plots represent the result of such computation for different values of the threshold. The blue curve represents the filling factors, i.e. the fraction of points used for computing the average with respect to the total number of points in the sample. The average of $D_i$ strongly increases when higher threshold are considered up to $J/J_{rms} = 10 $. 
Our results confirm that the exchange of energy is local, with larger values of $D_i$ localized in very small volume filling structures.
This evidence and the presence of non-Gaussian PDFs of dissipation proxies suggest that magnetic reconnection produces small scales current sheets which are site of strong events of energy exchange between fields and particles. Concisely stated, all these statistics indicate that dissipation in a reconnection event is intermittent. A similar conclusion was reached by \citet{wan2012intermittent} who examined the electron frame dissipation surrogate conditioned on magnitude of current density.

\begin{figure}
\includegraphics[width=\textwidth]{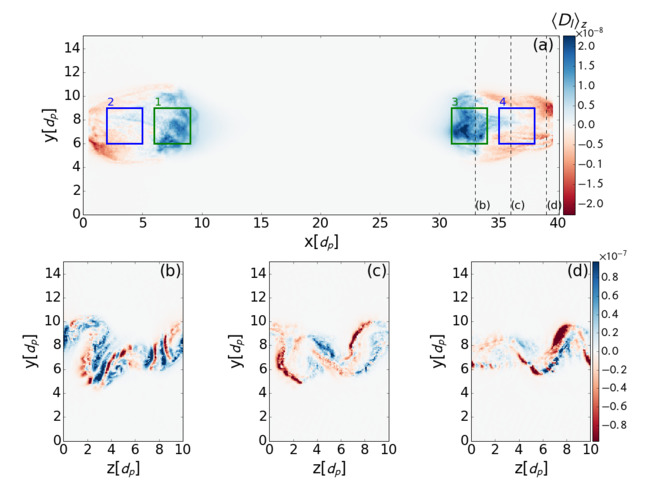}
\caption{Energy exchange $D_l = {\bf J} \cdot {\bf E}$ in the $xy$ plane averaged in the $z$ direction (a), and in the $yz$ plane at $x = 33\, d_i$ (b), 
        $x = 36\, d_i$ (c), $x = 39\, d_i$ (d). The x-line is located at $x =20 \, d_i$. The three boxes in panel (a) are the ones 
        used for the statical analysis presented in Figure~\ref{fig:Int_lb}-\ref{fig:Int_rb}.}
\label{fig:Dl_z}
\end{figure}

\begin{figure}
\includegraphics[width=\textwidth]{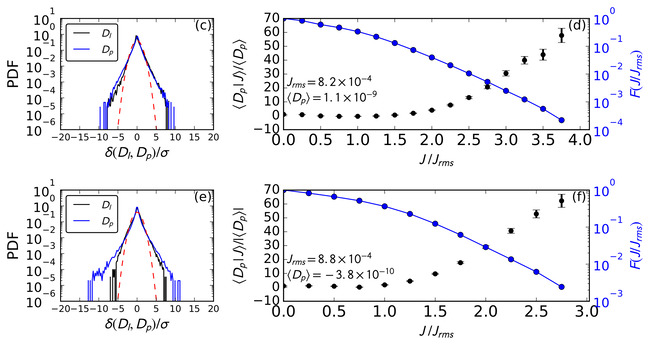}
\caption{PDFs of $D_l$ and $D_i$ in $BOX_1$ (c), $BOX_2$ (e). 
Conditioned average of $D_i$ and filling factors $F$ in $BOX_1$ (d), $BOX_2$ (f) (left outflow).}
\label{fig:Int_lb}
\end{figure}

\begin{figure}
\includegraphics[width=\textwidth]{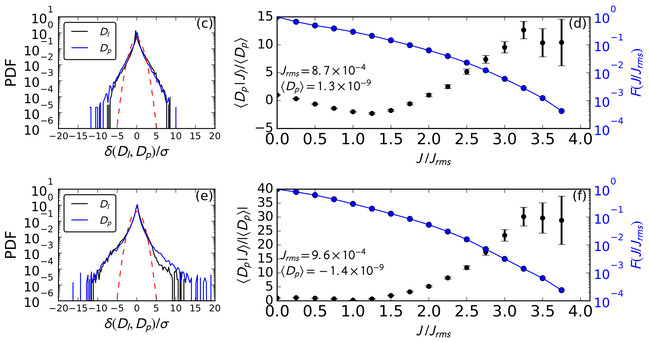}
\caption{PDFs of $D_l$ and $D_i$ in $BOX_3$ (c), $BOX_4$ (e). 
Conditioned average of $D_i$ and filling factors $F$ in $BOX_3$ (d), $BOX_4$ (f) (right outflow).}
\label{fig:Int_rb}
\end{figure}

Figure~\ref{fig:Int_lb}-\ref{fig:Int_rb} shows the statistics of dissipation proxies presented in Figure~\ref{fig:Int_fb} computed in sub-boxes located in the two reconnection outflows (see Figure~\ref{fig:Dl_z}). Non-Gaussian statistics and increasing conditioned average of dissipation proxies indicate that intermittent turbulence is at play in both reconnection outflows.

\section{Energy Exchanges in Reconnection outflows }
As shown in Figure~\ref{fig:nulls_slavik}, the region of the outflow is characterised by intense energy exchange ($\bfJ \cdot \bfE$). Recently the energy budget has been analysed in detail \citep{lapenta2016energy} and a large fraction of the energy is deposited as particle energization, while a significant fraction is also transported by the Poynting flux. 

\begin{figure}
  \centerline{\includegraphics[width=.7\columnwidth]{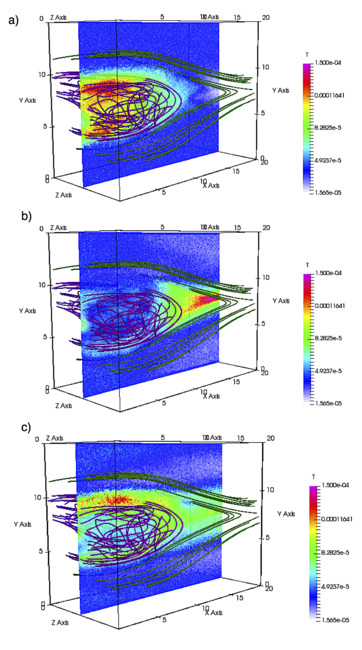}}
  \caption{Ion temperature in the outflow: from top to bottom: $T_x$ (a), $T_y$ (b) and $T_z$ (c). The left half  of the domain is reported at the final time along with selected field lines. }
\label{fig:Tion}
\end{figure}

Figure~\ref{fig:Tion} reports the ion temperature at the end of the run.  Ions are generally not magnetized in the reconnection region and projecting the pressure tensor in the parallel and perpendicular direction relative to the magnetic field is not productive. Ion energization in reconnection outflows and in reconnection fronts has been analysed in theory and in simulation \citep{aunai2011proton,pan2012adiabatic,birn2013particle,lapenta2016multiscale}. Complex processes are at play, requiring a full analysis of the phase space and of single particle trajectories to detect with accuracy the specific mechanisms accelerating the particles \citep{eastwood2015ion}.  

Figure~\ref{fig:Tion} reports the three different kinetic temperatures obtained from the pressure tensor: $T_{i} = P_{ii}/\rho_{i}$, for $i= x,\, y,\, z$. 
The primary region of reconnection tends to heat the ions primarily in the $y$ direction. This effect is due to the mixing of the two populations of ions coming from above and below from the inflow towards the reconnection region. In the outflow, instead, the plasma outflowing along the x-direction mixes with the plasma in the medium causing apparent heating in the x-direction \citep{aunai2011proton}. Heating in the $z$ direction is present both in the region of primary reconnection, where it is due to the acceleration of non-magnetised ions in the reconnection region due to the reconnection electric field \citep{moses1993plasma,divin2010model}, and in the region of the outflows, where it is a consequence of the instabilities in the outflows. These effects however should not be interpreted as heating in the meaning of increasing  thermodynamic temperature. The plasma is far from maxwellian and what appears as heating in the kinetic temperature (i.e. the second order moment of the distribution) is in reality the presence of multiple interpenetrating populations.

\begin{figure}
  \centerline{\includegraphics[width=\columnwidth]{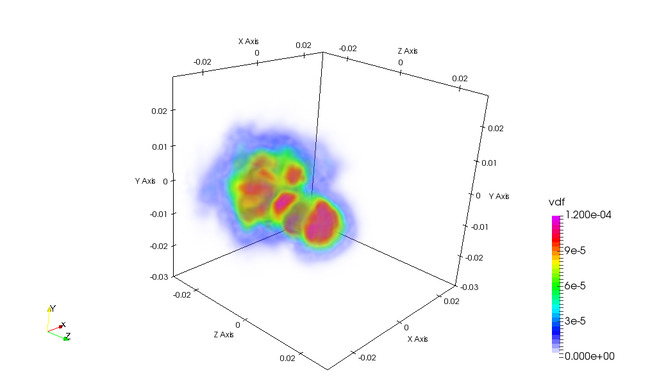}}
  \caption{Volume rendering of the ion velocity probability distribution $f_i(v_x,v_y,v_z)$ at the position $x/d_i=10.32$, $y/d_i=7.5$, $z/d_i=5$, obtained averaging over particles contained within a box centred at that location and with side $0.5d_i$. }
\label{fig:ion-distr}
\end{figure}

Figure~\ref{fig:ion-distr} shows a volume rendering of the full 3D velocity probability distribution for the ions. The distribution is anisotropic and contains multiple populations. When the second order moment is taken to measure a kinetic temperature, the result can be misleading because multiple beams, each with its own temperature, appear as a single plasma with a combined temperature much higher than that of the beams. However this is not a process of heating but one of bulk acceleration of ion populations. In a recent study, each ion component has been tracked back in time to its origin \citep{eastwood2015ion}. Each component originates from different regions and their trajectories brought them to the same location but with different speeds. 

\begin{figure}
  \centerline{\includegraphics[width=.7\columnwidth]{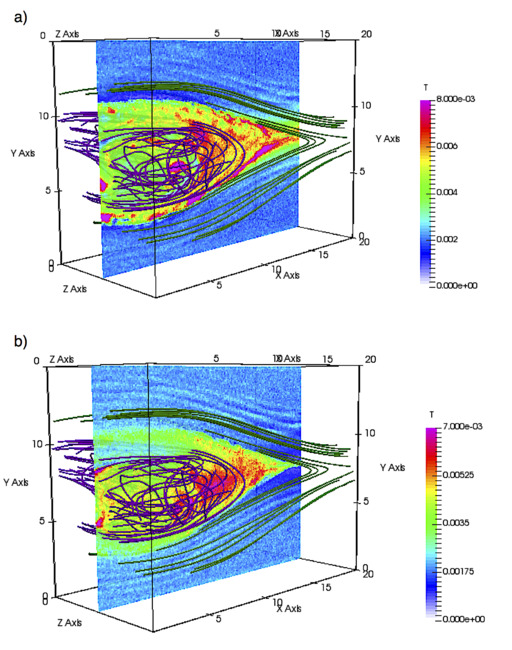}}
  \caption{Electron parallel (a) and perpendicular (b) temperature in the outflow. The left half  of the domain is reported at the final time along with selected field lines. }
\label{fig:Tele}
\end{figure}

Similarly, Figure~\ref{fig:Tele}  show the parallel and perpendicular electron temperature.  The electrons are mostly magnetised and it is more convenient to report the electron temperatures in magnetic coordinates rather than along geometrical axes. The region of primary reconnection causes parallel heating \citep{ricciacc}. The cause is the reconnection electric field that accelerates the electrons along the $z$ direction \citep{wan08d,divin2010model}: in this region the guide field is the only field present and the acceleration is parallel. The region of secondary instabilities in the outflow shows strong parallel and perpendicular energisation caused by the conversion of electromagnetic energy (i.e. $\bfJ_{e}\cdot \bfE$) \citep{lapenta2014electromagnetic,lapenta2016energy}.  

\begin{figure}
  \centerline{\includegraphics[width=\columnwidth]{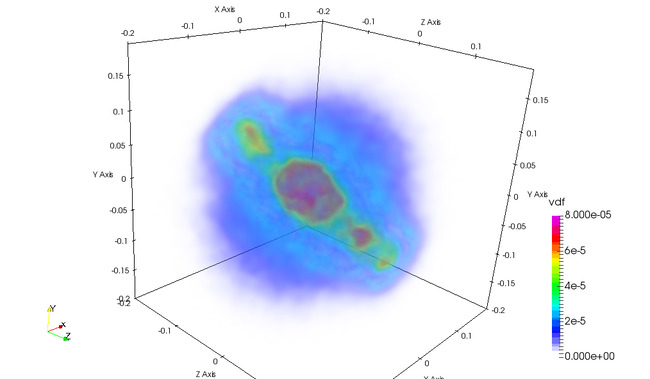}}
  \caption{Volume rendering of the electron velocity probability distribution $f_e(v_x,v_y,v_z)$ at the position $x/d_i=10.32$, $y/d_i=7.5$, $z/d_i=5$, obtained averaging over particles contained within a box centred at that location and with side $0.5d_i$.}
\label{fig:ele-distr}
\end{figure}

The electron distribution is typically far smoother than the ion distribution due to the higher thermal speed. However, in the region of the secondary front instability even the electron distribution becomes complex. Figure~\ref{fig:ele-distr} shows a volume rendering of the full 3D velocity probability distribution for the electrons computed as described above for the ions. On a large scale the distribution is bi-maxwellian with different parallel and perpendicular temperatures. Within it we can observe multiple electron populations caused by electron acceleration by the  electric field, directed primarily in the directions $z$ (reconnection electric field ) and $y$ (Hall electric field) \citep{wan08c}. Acceleration is not present in the $x$ direction where in fact there are no strong macroscopic electric fields.

The magnetic field lines show a chaotic behaviour: this condition makes it possible for particles moving along chaotic filed lines to access new acceleration regions of space \citep{dahlin2017role}, possibly encountering multiple reconnection sites and increasing  their energy in steps. 

\section{Conclusions and Future Directions}

The analysis of reconnection outflows in the present case of a weak guide field (1/10 of the main reconnecting field) show the development of an instability in the lower hybrid regime. In the present case, the instability has at least two components. The first, is due to the presence of density gradients formed in the pileup region where the outflow meets the ambient plasma. The second is the pre-existing velocity shears due to the differential velocity between the Harris plasma and the ambient plasma \citep{karimabadi2003ion,lapenta03,ricci-lhdi,riccietal04b}. 
The first instability leads to a Rayleigh-Taylor-type interchange instability in the lower hybrid range, while the latter leads to a kinking of the current layer. 

Both instabilities feed the onset of a turbulent cascade with the presence of coherent structures and intermittency. The outflows becomes host to secondary reconnection sites where the magnetic field topology becomes chaotic \citep{lapenta2015secondary}. 

We investigate here the effect of these processes on the energization of particles. The ions and the electrons are energized not only in the primary reconnection site but also, and in some cases predominantly, in the reconnection outflows. 
Particle energization can be linked to the electric fields operating on the particles. Electric fields do not heat particles in the statistical meaning of increasing their thermal spread, rather they coherently energise all particles, creating beams. Beams originating from different regions interact and  interpenetrate creating distribution functions with multiple populations. 

The end result is that the second order moment of the distribution is increased but the process cannot be interpreted as heating proper but rather as the presence of very non Maxwellian distributions with multiple beams. 


\section*{Acknowledgments}
The present work was supported by the 
Onderzoekfonds KU Leuven (Research Fund KU Leuven, GOA scheme and Space Weaves RUN project), by NASA's grant NNX08AO84G and by the US Air Force EOARD Award No. FA9550-14-1-0375. 
This research used resources of the National Energy Research
Scientific Computing Center, which is supported by the Office of
Science of the U.S. Department of Energy under Contract No. 
DE-AC02-05CH11231. Additional computing has been provided by NASA NAS and NCCS High Performance Computing, by the Flemish Supercomputing Center (VSC) and by PRACE Tier-0 allocations.




\end{document}